\documentclass[prd,twocolumn,amsfonts,amssymb]{revtex4}

\usepackage{graphicx}

\usepackage{bm}

\newcommand{\beq}{\begin{equation}}
\newcommand{\eeq}{\end{equation}}
\newcommand{\beqs}{\begin{eqnarray}}
\newcommand{\eeqs}{\end{eqnarray}}

\begin{document}

\title{Higher-Order Scheme-Independent Calculations of Physical Quantities 
in the Conformal Phase of a Gauge Theory}

\author{Thomas A. Ryttov$^a$ and Robert Shrock$^b$}

\affiliation{(a) \ CP$^3$-Origins and Danish Institute for Advanced Study \\
University of Southern Denmark, Campusvej 55, Odense, Denmark}

\affiliation{(b) \ C. N. Yang Institute for Theoretical Physics \\
Stony Brook University, Stony Brook, NY 11794, USA }

\begin{abstract}

  We consider an asymptotically free vectorial SU($N_c$) gauge theory with
  $N_f$ massless fermions in a representation $R$, having an infrared fixed
  point (IRFP) of the renormalization group at $\alpha_{IR}$ in the conformal
  non-Abelian Coulomb phase. The cases with $R$ equal to the fundamental,
  adjoint, and symmetric rank-2 tensor representation are considered. We
  present scheme-independent calculations of the anomalous dimension
  $\gamma_{\bar\psi\psi,IR}$ to $O(\Delta_f^4)$ and $\beta'_{IR}$ to
  $O(\Delta_f^5)$ at this IRFP, where $\Delta_f$ is an $N_f$-dependent
  expansion parameter.  Comparisons are made with conventional $n$-loop
  calculations and lattice measurements.  As a test of the accuracy of the
  $\Delta_f$ expansion, we calculate $\gamma_{\bar\psi\psi,IR}$ to
  $O(\Delta_f^3)$ in ${\cal N}=1$ SU($N_c$) supersymmetric quantum
  chromodynamics and find complete agreement, to this order, with the exactly
  known expression.  The $\Delta_f$ expansion also avoids a problem in which 
  an IRFP may not be manifest as an IR zero of a higher $n$-loop beta 
  function. 

\end{abstract}

\maketitle

A fundamental problem in quantum field theory concerns the properties at a
conformal fixed point of the renormalization group.  A specific question under
intensive current investigation concerns the properties of an asymptotically
free (AF) non-Abelian Yang-Mills vectorial gauge theory (in $d=4$ spacetime
dimensions) with a set of massless fermions at an IRFP of the renormalization
group in the Coulomb phase, where it exhibits scale and conformal invariance
\cite{fm,scalecon}.  Here we consider a theory of this type, with gauge group
$G = {\rm SU}(N_c)$ and $N_f$ massless fermions $\psi_j$, $1 \le j \le N_f$, in
a representation $R$, where $R$ is the fundamental ($F$), adjoint ($adj$), or
symmetric rank-2 tensor ($S$). The dependence of the gauge coupling $g=g(\mu)$
on the Euclidean momentum scale $\mu$ is described by the beta function, $\beta
= d\alpha/dt$, where $\alpha(\mu) = g(\mu)^2/(4\pi)$ and $dt=d\ln \mu$.  The
IRFP occurs at an IR zero of $\beta$ at $\alpha_{IR}$.  At this fixed point, an
operator ${\cal O}$ for a physical quantity exhibits scaling behavior with a
dimension $D_{\cal O}=D_{{\cal O},free}-\gamma_{\cal O}$, where $D_{{\cal
    O},free}$ is the free-field dimension and $\gamma_{\cal O}$ is the
anomalous dimension.

Two important quantities that characterize the properties at the IRFP
$\alpha_{IR}$ are $\gamma_{\bar\psi\psi}$ \cite{ops} and $\beta' \equiv
d\beta/d\alpha$, denoted $\gamma_{\bar\psi\psi,IR}$ and $\beta'_{IR}$.  Here,
$\beta'_{IR}$ is equivalent to the anomalous dimension of
$F_{a,\mu\nu}F_a^{\mu\nu}$, where $F_a^{\mu\nu}$ is the (rescaled)
field-strength tensor \cite{gammaff}.  As physical quantities,
$\gamma_{\bar\psi\psi,IR}$ and $\beta'_{IR}$ are scheme-independent (SI)
\cite{gross75}. However, conventional series expansions of these quantities in
powers of $\alpha$, calculated to a finite order, do not maintain this
scheme-independence beyond the lowest orders.  Clearly, it is very valuable to
calculate and analyze series expansions for $\gamma_{\bar\psi\psi,IR}$ and
$\beta'_{IR}$ that are scheme-independent at each order. Some early work was in
\cite{bz,grunberg92}.  A natural expansion variable is
\beq
\Delta_f = N_u-N_f \ , 
\label{deltaf}
\eeq
where, for a given $N_c$ and $R$, $N_u$ is the upper ($u$) limit to $N_f$
allowed by asymptotic freedom.  Scheme-independent series expansions of 
$\gamma_{\bar\psi\psi,IR}$ and $\beta'_{IR}$ are \cite{gtr}
\beq
\gamma_{\bar\psi\psi,IR} = \sum_{j=1}^\infty \kappa_j \, \Delta_f^j 
\label{gamma_ir_Deltaseries}
\eeq
and \cite{dex} 
\beq
\beta'_{IR} = \sum_{j=1}^\infty d_j \, \Delta_f^j \ ,
\label{betaprime_ir_Deltaseries}
\eeq
where $d_1=0$ for all $G$ and $R$.  For general $G$ and $R$, the $\kappa_j$
were calculated to order $j=3$ in \cite{gtr} and the $d_j$ to order $j=4$ in
\cite{dex}, and for $G={\rm SU}(3)$ and $R=F$, $\kappa_4$ was computed in
\cite{gsi} and $d_5$ in \cite{dex}.  

Here we report our calculations of these scheme-independent expansions of
$\gamma_{\bar\psi\psi,IR}$ and $\beta'_{IR}$ to the highest orders yet
achieved, presenting $\kappa_4$ and $d_5$ for an asymptotically free SU($N_c$)
gauge theory with a conformal IR fixed point, for $R=F, \ adj, \ S$.  We also
report our calculation of $\kappa_3$ for supersymmetric quantum chromodynamics
(SQCD).  We believe that our new results are a substantial advance in the
knowledge of conformal field theory.  Our results have the advantage of scheme
independence at each order in $\Delta_f$, in contrast to scheme-dependent (SD)
series expansions of $\gamma_{\bar\psi\psi,IR}$ and $\beta'_{IR}$ in powers of
$\alpha$ \cite{gk}-\cite{flir} and they complement other approaches to
understanding conformal and superconformal field theory, such as the bootstrap
\cite{cft_bootstrap} and lattice simulations \cite{lgtreviews}.

The conventional power-series expansions of
$\beta$ and $\gamma_{\bar\psi\psi}$ are
\beq
\beta = -2\alpha \sum_{\ell=1}^\infty b_\ell \,
\Big (\frac{\alpha}{4\pi} \Big )^\ell \
\label{beta}
\eeq
and
\beq
\gamma_{\bar\psi\psi} = \sum_{\ell=1}^\infty
c_\ell \Big ( \frac{\alpha}{4\pi} \Big )^\ell \ ,
\label{gamma}
\eeq
where $b_\ell$ and $c_\ell$ are the $\ell$-loop coefficients;
$b_1$ \cite{b1}, $b_2$ \cite{b2}, and
$c_1=6C_f$ are scheme-independent, while the $b_\ell$ with
$\ell \ge 3$ and the $c_\ell$ with $\ell \ge 2$ are
scheme-dependent, i.e. they depend on the scheme used for
regularization and renormalization \cite{gross75}. We denote the 
$n$-loop ($n \ell$) $\beta$ and $\gamma_{\bar\psi\psi}$ as 
$\beta_{n\ell}$ and $\gamma_{\bar\psi\psi,n\ell}$ and the IR zero of 
$\beta_{n\ell}$ as $\alpha_{n\ell}$. 

The calculation of $\kappa_j$ requires, as inputs, the values of the $b_\ell$
for $1 \le \ell \le j+1$ and the $c_\ell$ for $1 \le \ell \le j$.  The
calculation of $d_j$ requires, as inputs, the values of the $b_\ell$ for $1 \le
\ell \le j$. Thus, importantly, $\kappa_j$ does not receive any corrections
from $b_\ell$ with $\ell > j+1$ or $c_\ell$ with $\ell > j$, and similarly,
$d_j$ does not receive any corrections from any $b_\ell$ with $\ell > j$. 

The coefficients $\kappa_j$ were calculated in \cite{gtr} for an (AF vectorial)
supersymmetric gauge theory (SGT) with gauge group $G$ and $N_f$ pairs of
chiral superfields in the $R$ and $\bar R$ representation, for $j=1, \
2$. Complete agreement was found, to the order calculated, with the exactly
known result in the conformal non-Abelian Coulomb phase (NACP)
\cite{nsvz,seiberg,casimir}
\beq
\gamma_{IR,SGT} = \frac{\frac{2T_f}{3C_A}\Delta_f}
{1-\frac{2T_f}{3C_A}\Delta_f } \ . 
\label{gamma_ir_sgt}
\eeq
In this theory, $N_u=3C_A/(2T_f)$, and the conformal NACP is the interval
$N_\ell < N_f < N_u$, where $N_\ell = N_u/2$, so that
$\Delta_f$ varies from 0 to a maximum of $(\Delta_f)_{max}=3C_A/(4T_f)$ in the
NACP \cite{nfintegral}.  Hence, $\gamma_{IR,SGT}$ increases monotonically from
0 to 1 as $N_f$ decreases from $N_u$ to $N_\ell$, saturating the upper bound
$\gamma_{\bar\psi\psi,IR,SGT} < 1$ from conformal invariance in this SGT
\cite{gammabound}. 

As a test of the accuracy of the $\Delta_f$ expansion, we have now calculated
$\kappa_3$ for SQCD with $R=F$, using inputs from \cite{b4susy}. We find
$\kappa_3 = 1/(3N_c)^3$, in perfect agreement, to this order, with the exact
result, Eq. (\ref{gamma_ir_sgt}).  This agreement explicitly illustrates the
scheme independence of the $\kappa_j$, since our calculations in \cite{gtr} and
here used inputs computed in the $\overline{DR}$ scheme, while
(\ref{gamma_ir_sgt}) was derived in the NSVZ scheme \cite{nsvz}.  Our new
result has a far-reaching implication: it strongly suggests that $\kappa_j =
[2T_f/(3C_A)]^j$ for all $j$, so that the expansion
(\ref{gamma_ir_Deltaseries}) for this supersymmetric gauge theory, calculated
to order $O(\Delta_f^p)$, agrees with the exact result to the given order for
all $p$.

Because of electric-magnetic duality \cite{seiberg}, as $N_f \to
N_\ell$ in the NACP, the physics is described by a magnetic theory with
coupling strength going to zero, or equivalently, by an electric theory with
divergent $\alpha_{IR}$. Hence, another important finding here is that the
complete agreement that we obtain in SQCD to $O(\Delta_f^3)$ between Eq.
(\ref{gamma_ir_Deltaseries}) and the exact Eq. (\ref{gamma_ir_sgt}) holds for
arbitrarily strong $\alpha_{IR}$.  Even apart from the issue of scheme
dependence in Eq. (\ref{gamma}), this agreement could not be achieved with the
conventional expansion (\ref{gamma}) of $\gamma_{\bar\psi\psi,IR}$ in powers of
$\alpha$.  

The $\Delta_f$ expansion also avoids a problem in which an IRFP may not be
manifest as a physical IR zero of the $n$-loop beta function for some $n$. 
Indeed, although $\beta_{n\ell}$ has a physical $\alpha_{IR,n\ell}$ in SQCD for
$n=2, \ 3$ loops \cite{bfs}, we have analyzed $\beta_{4\ell}$ (in the
$\overline{DR}$ scheme), and we find that for a range of $N_f$ in the NACP, it
does not exhibit a physical $\alpha_{IR,4\ell}$. This is analogous to the
situation that we found for $\alpha_{IR,5\ell}$ in the non-supersymmetric gauge
theory \cite{flir}.  In both cases, the $\Delta_f$ expansions
(\ref{gamma_ir_Deltaseries}) and (\ref{betaprime_ir_Deltaseries}) circumvent
this problem of a possible unphysical $\alpha_{IR,n\ell}$ that one may
encounter in using the convention expansions (\ref{beta}) and (\ref{gamma}).

We next present our results for $\kappa_4$ and $d_5$ for a (non-supersymmetric)
SU($N_c$) gauge theory, making use of the impressive recent computation of
$b_5$ in \cite{b5}. (We have actually calculated $\kappa_4$ and $d_5$ for
general $G$ and $R$ \cite{dex2}, but only present results here for $R=F,\ adj,
\ S$.)  The two-loop beta function has an IR zero (IRZ) in the interval
$I_{IRZ}: \ N_\ell < N_f < N_u$, with upper and lower ($\ell$) ends at $N_u
=11N_c/(4T_f)$ and $N_\ell =17C_A^2/[2T_f(5C_A+3C_f)]$ \cite{nfintegral}. The
non-Abelian Coulomb phase extends downward in $I_{IRZ}$ from $N_u$ to a lower
value denoted $N_{f,cr}$ \cite{nfcr}.  Since chiral symmetry is exact in the
NACP, one can classify the bilinear fermion operators according to their flavor
transformation properties.  These operators include the flavor-singlet
$\bar\psi\psi$ and the flavor-adjoint $\bar\psi T_a \psi$, where $T_a$ is a
generator of SU($N_f$). These have the same anomalous dimension
\cite{gracey_tensor}, which we write simply as $\gamma_{\bar\psi\psi}$.  For
general $G$ and $R$, the coefficients $b_\ell$ were computed up to loop order
$\ell=4$ \cite{b4} (checked in \cite{b4c}) and the $c_\ell$ also up to loop
order $\ell=4$ \cite{c4}, in the widely used $\overline{\rm MS}$ scheme
\cite{msbar}. These results were used in \cite{gtr} to calculate the $\kappa_j$
to order $j=3$ and in \cite{dex} to calculate $d_j$ to order $j=4$. For $N_c=3$
and $R=F$, $b_5$ was computed in \cite{b5su3}, and this was used to calculate
$\kappa_4$ in \cite{gsi} and $d_5$ in \cite{dex} for this case (see also
\cite{flir}).
 
We first report our results for $\kappa_4$ 
and $d_5$ for $R=F$, using $b_5$ from \cite{b5}.  We denote the Riemann 
zeta function as $\zeta_s =\sum_{n=1}^\infty n^{-s}$.  We obtain 
\begin{widetext}
\beqs
\kappa_{4,F} &=&
\frac{4(N_c^2-1)}{3^4N_c^4(25N_c^2-11)^7} \,
\bigg [ \Big (263345440N_c^{12} - 673169750N_c^{10} + 256923326N_c^8 \cr\cr
&-& 290027700N_c^6 + 557945201N_c^4 - 208345544N_c^2 + 6644352 \Big ) \cr\cr
&+& 384(25N_c^2-11)\Big ( 4400N_c^{10}-123201N_c^8+480349N_c^6-486126N_c^4
+84051N_c^2+1089 \Big )\zeta_3 \cr\cr
&+& 211200N_c^2(25N_c^2-11)^2(N_c^6+3N_c^4-16N_c^2+22)\zeta_5 \ \bigg ]
\eeqs
and 
\beqs
d_{5,F} & = & \frac{2^5}{3^6N_c^3 (25N_c^2-11)^7} \, \bigg [
N_c^{12}\Big (-298194551 - 423300000\zeta_3 + 528000000\zeta_5 \Big ) \cr\cr
&+& N_c^{10}\Big (414681770+1541114400\zeta_3 - 821040000\zeta_5 \Big )
     + N_c^8\Big (80227411-4170620256\zeta_3+2052652800\zeta_5 \Big ) \cr\cr
&+& N_c^6\Big (210598856+5101712352\zeta_3-4268183040\zeta_5 \Big )
+ N_c^4 \Big (-442678324-2250221952\zeta_3+2744628480\zeta_5\Big ) \cr\cr
&+& N_c^2 \Big (129261880+304571520\zeta_3-534103680\zeta_5 \Big ) +
3716152+1022208\zeta_3 \bigg ] \ ,
\label{d5f}
\eeqs
\end{widetext}
where the simple factorizations of the denominators have been indicated.  For
this $R=F$ case, we find that $\kappa_4 > 0$, as was also true of $\kappa_j$
with $1 \le j \le 3$ (indeed, $\kappa_1$ and $\kappa_2$ are manifestly positive
for any $G$ and $R$).  We also find the same positivity results for $R=adj$ and
$R=S$. The property that for all of these representations 
$R$, $\kappa_j > 0$ for $1 \le j \le
4$ and for all $N_c$ implies two important monotonicity results. First, for
these $R$, and with a fixed $p$ in the interval $1 \le p \le 4$,
$\gamma_{\bar\psi\psi,IR,\Delta_f^p}$ is a monotonically increasing function of
$\Delta_f$ for $N_f \in I_{IRZ}$. Second, for these $R$, and with a fixed $N_f
\in I_{IRZ}$, $\gamma_{\bar\psi\psi,IR,\Delta_f^p}$ is a monotonically
increasing function of $p$ in the range $1 \le p \le 4$.  
In addition to the manifestly positive $\kappa_1$ and $\kappa_2$, a plausible 
conjecture is that, for these $R$, $\kappa_j > 0$ for all $j \ge 3$. 
Note that the exact result (\ref{gamma_ir_sgt}) for the supersymmetric
gauge theory shows that in that theory, $\kappa_j > 0$ for all $j$ and for any
$G$ and $R$.

\begin{figure}
  \begin{center}
    \includegraphics[height=6cm]{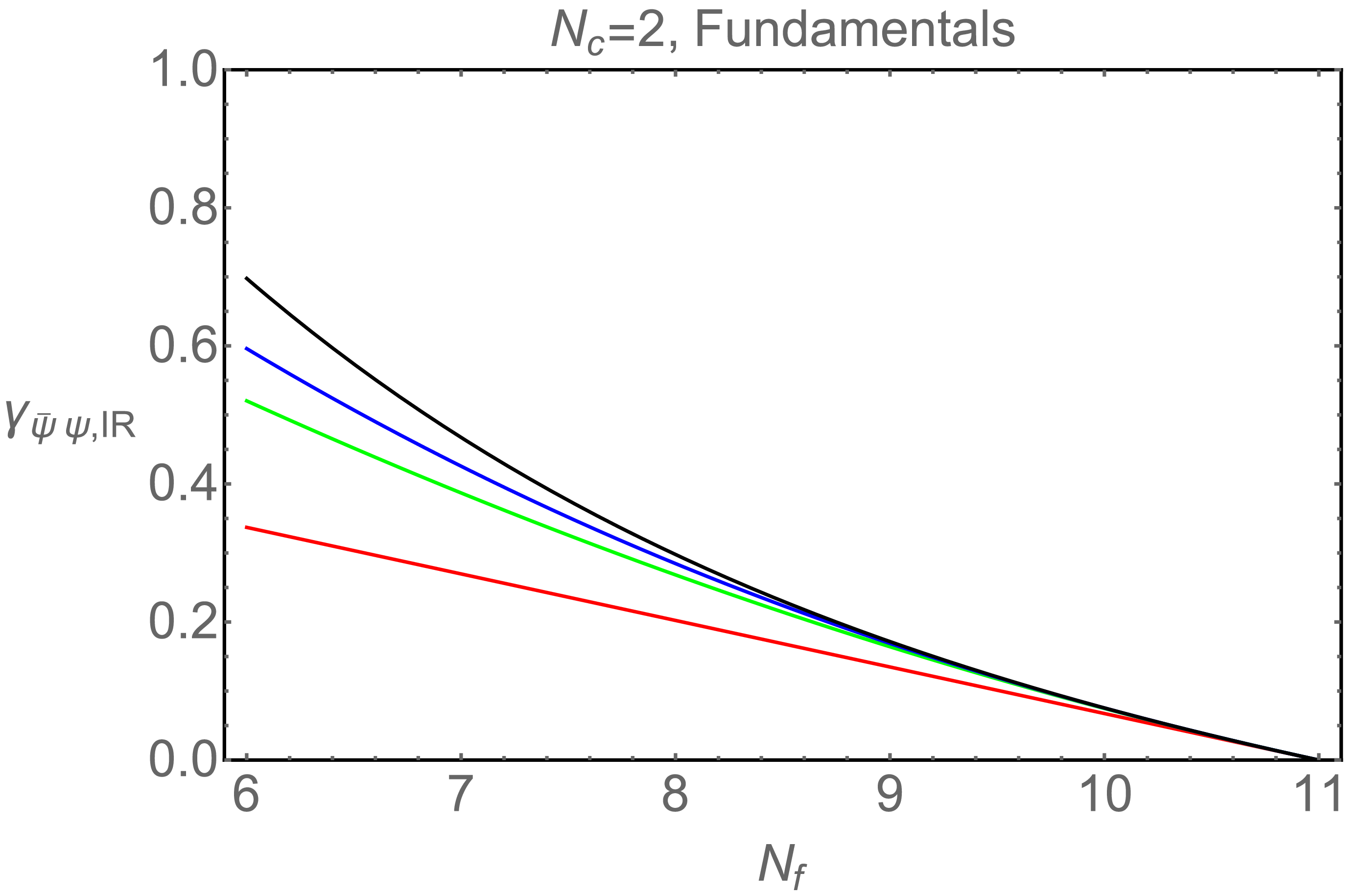}
  \end{center}
\caption{Plot of $\gamma_{\bar\psi\psi,IR,\Delta_f^p}$ for $R=F$, $N_c=2$, and
$1 \le p \le 4$ as a function of $N_f \in I_{IRZ}$.  From bottom to top,
the curves (with colors online) refer to
$\gamma_{\bar\psi\psi,IR,\Delta_f}$ (red),
$\gamma_{\bar\psi\psi,IR,\Delta_f^2}$ (green),
$\gamma_{\bar\psi\psi,IR,\Delta_f^3}$ (blue), and
$\gamma_{\bar\psi\psi,IR,\Delta_f^4}$ (black).}
\label{gammaNc2fund_plot}
\end{figure}

\begin{figure}
  \begin{center}
    \includegraphics[height=6cm]{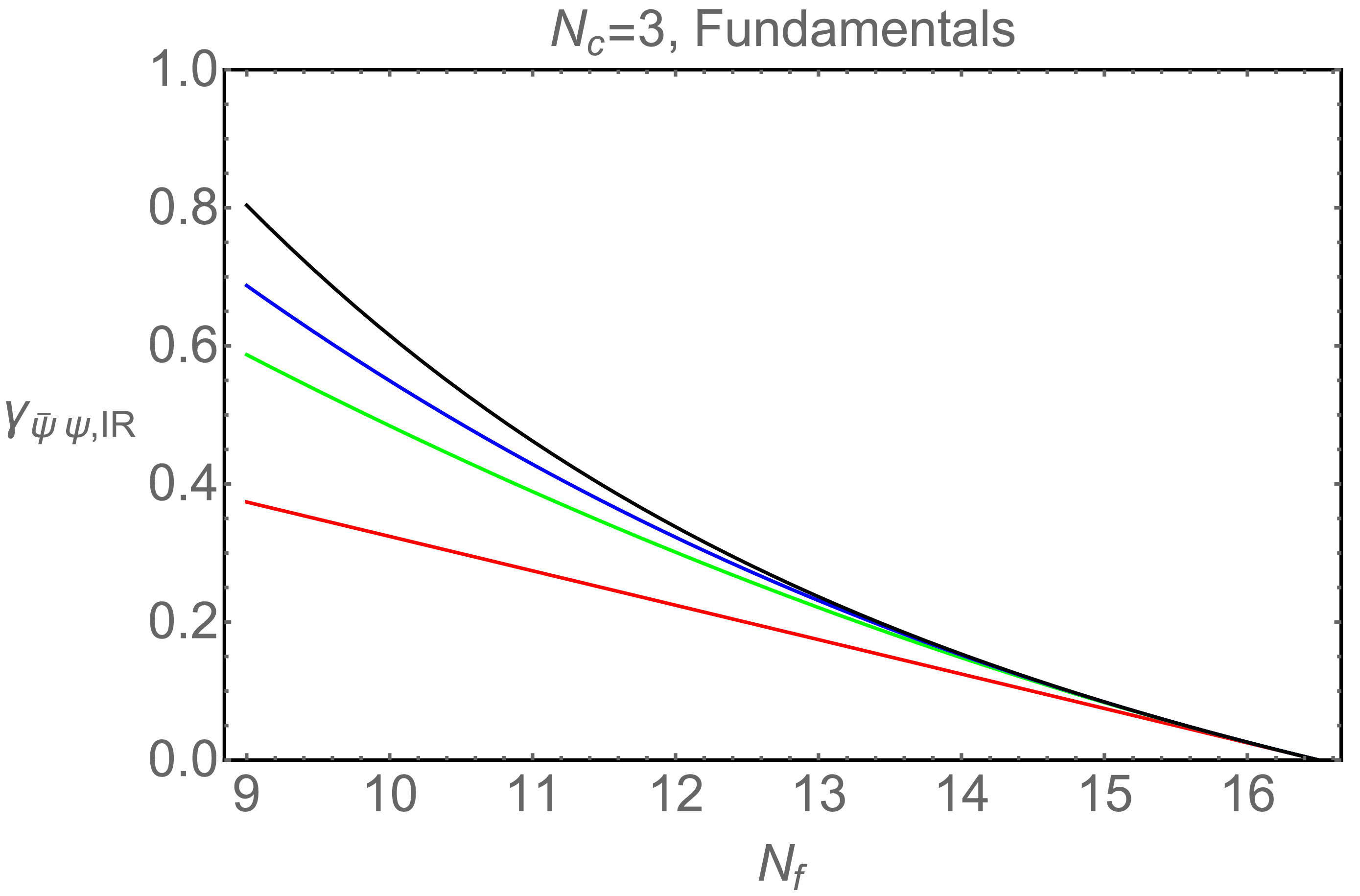}
  \end{center}
\caption{Plot of $\gamma_{\bar\psi\psi,IR,\Delta_f^p}$ for $R=F$, $N_c=3$, and
$1 \le p \le 4$ as a function of $N_f \in I_{IRZ}$.  From bottom to top,
the curves (with colors online) refer to
$\gamma_{\bar\psi\psi,IR,\Delta_f}$ (red),
$\gamma_{\bar\psi\psi,IR,\Delta_f^2}$ (green),
$\gamma_{\bar\psi\psi,IR,\Delta_f^3}$ (blue), and
$\gamma_{\bar\psi\psi,IR,\Delta_f^4}$ (black).}
\label{gammaNc3fund_plot}
\end{figure}

In Figs. \ref{gammaNc2fund_plot} and \ref{gammaNc3fund_plot} we plot
$\gamma_{IR,\Delta_f^p}$ for $R=F$, $N_c=2, \ 3$ and $1 \le p \le 4$.  In Table
\ref{gamma_values} we list values of these $\gamma_{IR,\Delta_f^p}$ 
\ \cite{convergence}.  These all
satisfy the upper bound $\gamma_{IR} < 2$ from conformal invariance
\cite{gammabound}.  Below, we will often omit the $\bar\psi\psi$
subscript, writing $\gamma_{\bar\psi\psi,IR} \equiv \gamma_{IR}$ and
$\gamma_{\bar\psi\psi,IR,\Delta_f^p} \equiv \gamma_{IR,\Delta_f^p}$.

\begin{table}
\caption{\footnotesize{Values of the scheme-independent anomalous dimension    
$\gamma_{IR,\Delta_f^p}$ with $1 \le p \le 4$ for $R=F$ and $N_c=2, \ 3$.}}
\begin{center}
\begin{tabular}{|c|c|c|c|c|c|} \hline\hline
$N_c$ & $N_f$ &
$\gamma_{IR,\Delta_f}$  &
$\gamma_{IR,\Delta_f^2}$ &
$\gamma_{IR,\Delta_f^3}$ &
$\gamma_{IR,\Delta_f^4}$
\\ \hline
2& 6& 0.337  & 0.520   & 0.596  & 0.698   \\
2& 7& 0.270  & 0.387   & 0.426  & 0.467   \\
2& 8& 0.202  & 0.268   & 0.285  & 0.298   \\
2& 9& 0.135  & 0.164   & 0.169  & 0.172   \\
2&10& 0.0674 & 0.07475 & 0.07535& 0.0755  \\
\hline
3& 9& 0.374  & 0.587  & 0.687  & 0.804    \\
3&10& 0.324  & 0.484  & 0.549  & 0.615    \\
3&11& 0.274  & 0.389  & 0.428  & 0.462    \\
3&12& 0.224  & 0.301  & 0.323  & 0.338    \\
3&13& 0.174  & 0.221  & 0.231  & 0.237    \\
3&14& 0.125  & 0.148  & 0.152  & 0.153    \\
3&15& 0.0748 & 0.0833 & 0.0841 & 0.0843   \\
3&16& 0.0249 & 0.0259 & 0.0259 & 0.0259   \\
\hline\hline
\end{tabular}
\end{center}
\label{gamma_values}
\end{table}

For this $R=F$ case we first remark on the comparison of
$\gamma_{IR,\Delta_f^4}$ with calculations of $\gamma_{IR,n\ell}$ from analyses
of power series in $\alpha$, which were performed to $n=4$ loop level in
\cite{gk}-\cite{ps} using $b_\ell$ and $c_\ell$ in the $\overline{\rm MS}$
scheme (with studies of scheme dependence in \cite{sch}) and
extended to $n=5$ loop level for $N_c=3$ in \cite{flir}.  We have noted that 
$\beta_{5\ell}$ does not have a physical $\alpha_{IR,5\ell}$ for $N_f$ in 
the lower part of the interval $I_{IRZ}$ \cite{flir}.
Although we were able to surmount this problem via Pad\'e approximants in 
\cite{flir}, these are still scheme-dependent, while the $\Delta_f$ expansion
has the advantage of being scheme-independent.  In general, we find that for a
given $N_c$ and $N_f$, the value of $\gamma_{IR,\Delta_f^p}$ that we calculate
to highest order, namely $p=4$, is somewhat larger than $\gamma_{IR,n\ell}$ 
calculated to its highest order \cite{bvh,gsi}. For example, for $N_c=3$,
$N_f=12$, $\gamma_{IR,4\ell}=0.253$, $\gamma_{IR,5\ell} \simeq 0.255$ (using a
value of $\alpha_{IR,5\ell}$ from a Pad\'e approximant \cite{gsi,flir}), while
$\gamma_{IR,\Delta_f^4}=0.338$ and an extrapolation yields the estimate
0.400(5) for $\gamma_{IR} = \lim_{p \to \infty} \gamma_{IR,\Delta_f^p}$
\cite{gsi}.  Similarly, for $N_c=2$ and $N_f=8$, $\gamma_{IR,4\ell}=0.204$, 
while $\gamma_{IR,\Delta_f^4}=0.298$; and for $N_c=4$,
$N_f=16$, $\gamma_{IR,4\ell}=0.269$, while $\gamma_{IR,\Delta_f^4}=0.352$.

We next compare our new results with lattice measurements, restricting to cases
where the lattice studies are consistent with the theories being IR-conformal
\cite{nfcr,lgtreviews}. For $N_c=3$, we compared our calculations of
$\gamma_{IR,\Delta_f^4}$ with lattice measurements for $N_f=12$ in \cite{gsi},
finding general consistency with the range of lattice results, although our
$\gamma_{IR,\Delta_f^4}$ and extrapolation to the exact $\gamma_{IR}$ were
higher than some of the lattice values. We also found consistency for the cases
$N_f=10$ and $N_f=8$ \cite{gsi}. Here, we compare with lattice results for
$\gamma_{IR}$ in the case $N_c=2$, $N_f=8$. (It is not clear from lattice
studies if the SU(2), $R=F$, $N_f=6$ theory has a conformal IRFP or not
\cite{nfcr,lgtreviews,su2nf6}.)  Following lattice studies of the SU(2), $R=F$,
$N_f=8$ theory by several groups \cite{lgtreviews,su2nf8}, a recent measurement
is $\gamma_{IR}=0.15 \pm 0.02 \equiv 0.15(2)$ \cite{tuominen2017}. Our value
$\gamma_{IR,\Delta_f^4}=0.298$ is somewhat higher than this lattice result.

\begin{figure}
  \begin{center}
    \includegraphics[height=6cm]{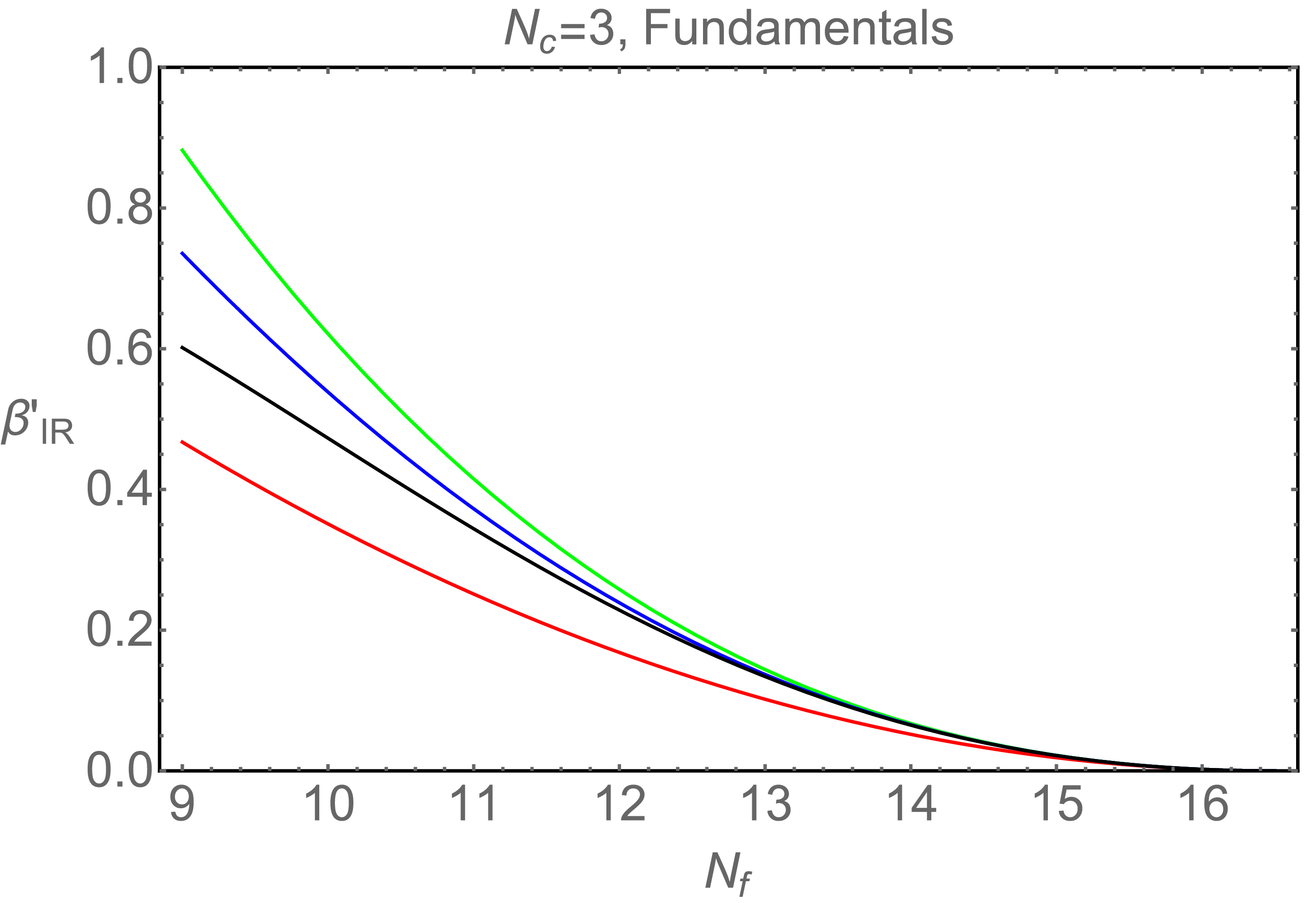}
  \end{center}
\caption{Plot of $\beta'_{IR,\Delta_f^p}$ for $R=F$, $N_c=3$, and
$2 \le p \le 4$ as a function of $N_f \in I_{IRZ}$.  From bottom to top,
the curves (with colors online) refer to
$\beta'_{IR,\Delta_f^2}$ (red),
$\beta'_{IR,\Delta_f^3}$ (green),
$\beta'_{IR,\Delta_f^4}$ (blue), 
$\beta'_{\bar\psi\psi,IR,\Delta_f^5}$ (black).}
\label{betaprimeNc3fund_plot}
\end{figure}

\begin{table}
\caption{\footnotesize{ Scheme-independent values of                           
$\beta'_{IR,\Delta_f^p}$ with $2 \le p \le 4$ for $R=F$, $N_c=2, \ 3$ 
as functions of $N_f$ in the respective intervals $I_{IRZ}$. 
The notation $a$e-$n$ means $a \times 10^{-n}$.}}
\begin{center}
\begin{tabular}{|c|c|c|c|c|c|} \hline\hline
$N_c$ & $N_f$
    & $\beta'_{IR,\Delta_f^2}$
    & $\beta'_{IR,\Delta_f^3}$
    & $\beta'_{IR,\Delta_f^4}$
    & $\beta'_{IR,\Delta_f^5}$
\\ \hline
 2 & 6 & 0.499   & 0.957   & 0.734  & 0.6515 \\
 2 & 7 & 0.320   & 0.554   & 0.463  & 0.436  \\
 2 & 8 & 0.180   & 0.279   & 0.250  & 0.243  \\
 2 & 9 & 0.0799  & 0.109   & 0.1035 & 0.103  \\
 2 &10 & 0.0200  & 0.0236  & 0.0233 & 0.0233 \\
 \hline
 3 & 9 & 0.467   & 0.882   & 0.7355  & 0.602 \\
 3 &10 & 0.351   & 0.621   & 0.538   & 0.473 \\
 3 &11 & 0.251   & 0.415   & 0.3725  & 0.344 \\
 3 &12 & 0.168   & 0.258   & 0.239   & 0.228 \\
 3 &13 & 0.102   & 0.144   & 0.137   & 0.134 \\
 3 &14 & 0.0519  & 0.0673  & 0.0655  & 0.0649 \\
 3 &15 & 0.0187  & 0.0220  & 0.0218  & 0.0217 \\
 3 &16 & 2.08e-3 & 2.20e-3 & 2.20e-3& 2.20e-3 \\
\hline\hline
\end{tabular}
\end{center}
\label{betaprime_values}
\end{table}

We proceed to discuss $d_5$ for $R=F$. In Fig. \ref{betaprimeNc3fund_plot} we
plot $\beta'_{IR,\Delta_f^p}$ for $R=F$, $N_c=3$, and $2 \le p \le 5$. In Table
\ref{betaprime_values} we list values of $\beta'_{IR,\Delta_f^p}$ for $R=F$,
$N_c=2, \ 3$ and $2 \le p \le 5$.  For $R=F$ and general $N_c$, $d_2$ and $d_3$
are positive, while $d_4$ and $d_5$ are negative. For the case SU(3), $N_f=12$,
we get $\beta'_{IR,\Delta_f^5}=0.228$.  The conventional $n$-loop calculation
yielded $\beta'_{IR,3\ell}=0.2955$ and $\beta'_{IR,4\ell}=0.282$ \cite{bc}, so
$\beta'_{IR,\Delta_f^5}$ is slightly smaller than $\beta'_{IR,4\ell}$.  A
recent lattice measurement yields $\beta'_{IR} = 0.26(2)$ \cite{hs2016},
consistent with both our $\beta'_{IR,\Delta_f^5}$ and $\beta'_{IR,4\ell}$.

We next discuss the case $R=adj$, for which 
$N_u=11/4$ and $N_\ell=17/16$, so $I_{IRZ}$ includes the single integer value
$N_f=2$ (whence $\Delta_f = N_u-2=3/4$). Results for this case were given for 
$\kappa_p$ with $1\le p \le 3$ in \cite{gtr} and for $d_p$ with $1 \le p \le 4$
in \cite{dex}. Here we find
\beqs
& & \kappa_{4,adj} = \frac{53389393}{2^7 \cdot 3^{14}} +
\frac{368}{3^{10}}\zeta_3 \cr\cr
& + &
\bigg ( -\frac{2170}{3^{10}} + \frac{33952}{3^{11}}\zeta_3 \bigg )N_c^{-2}
\cr\cr
& = & 0.0946976 + 0.193637 N_c^{-2}
\label{kappa4_adj}
\eeqs
and
\beqs
d_{5,adj} & = & -\frac{7141205}{2^3 \cdot 3^{16}}
+ \frac{5504}{3^{12}}\zeta_3 \cr\cr
&-& \bigg ( \frac{30928}{3^{14}} +
\frac{465152}{3^{13}}\zeta_3 \bigg )N_c^{-2}  \cr\cr
& = & -(0.828739 \times 10^{-2}) - 0.357173 N_c^{-2} \ . \cr\cr
& & 
\label{d5_adj}
\eeqs
We remark on the SU(2), $N_f=2$, $R=adj$ theory, which has been of interest
\cite{sannino_su2adj}. Extensive lattice studies of this theory have been
performed and are consistent with IR conformality \cite{lgtreviews}.
We get $\beta'_{IR,\Delta_f^5}=0.147$; and 
$\gamma_{IR,\Delta_f^2}=0.465$, $\gamma_{IR,\Delta_f^3}=0.511$, and 
$\gamma_{IR,\Delta_f^4}=0.556$. These $\gamma_{IR,\Delta_f^p}$ values 
are close to our $n$-loop calculations in \cite{bvh} for this theory, namely
$\gamma_{IR,3\ell}=0.543$, $\gamma_{IR,4\ell}=0.500$. 
Lattice measurements of this theory have yielded a wide range of
values of $\gamma_{IR}$ including, 0.49(13) \cite{catterall2010}, 0.22(6)
\cite{deldebbio2010}, 0.31(6) \cite{degrand2011}, 0.17(5) \cite{lsd2011}, 
0.20(3) \cite{tuominen2015},
0.50(26) \cite{giedt2016}, and 0.15(2) \cite{tuominen2017} (see references for
details of uncertainty estimates).

Finally, we discuss the case $R=S$. For SU(2), $S=adj$, already discussed
above.  For SU(3), we focus on the $N_f=2$ theory, for which we find
$\beta'_{IR,\Delta_f^5}=0.333$; and $\gamma_{IR,\Delta_f^2}=0.789$,
$\gamma_{IR,\Delta_f^3}=0.960$, and $\gamma_{IR,\Delta_f^4}=1.132$ \
\cite{convergence}.  For comparison, our $n$-loop results from \cite{bvh} for
this case are $\gamma_{IR,3\ell}=0.500$ and $\gamma_{IR,4\ell}=0.470$. Lattice
studies of this theory include one that concludes that it is IR-conformal and
gets $\gamma_{IR} < 0.45$ \cite{degrand_sextet} and another that concludes that
it is not IR-conformal and gets an effective $\gamma_{IR} \simeq 1$
\cite{kuti_sextet}.

In summary, we have presented calculations of $\gamma_{\bar\psi\psi,IR}$ and
$\beta'_{IR}$ at a conformal IR fixed point of an asymptotically free gauge
theory with fermions, to the highest orders yet achieved. We
believe that these results are of fundamental value for the understanding of
conformal field theory, especially because they are scheme-independent.

This research was supported in part by the Danish National
Research Foundation grant DNRF90 to CP$^3$-Origins at SDU (T.A.R.) and by 
the U.S. NSF Grant NSF-PHY-16-1620628 (R.S.).


\begin{thebibliography}{99}
\bibitem{fm}
%
The assumption of massless fermions incurs no loss of generality, since 
if a fermion had a nonzero mass $m_0$, it would be integrated out of
the effective field theory at scales $\mu < m_0$, and hence would not
affect the IR limit $\mu \to 0$.
\bibitem{casimir}
%
$C_A$ and $C_f$ are the quadratic Casimir invariants for the adjoint
representation and the fermion representation $R$, and $T_f$ is the trace
invariant. We use the standard normalizations for these, so that for
$G={\rm SU}(N_c)$, $C_A=N_c^2-1$ and for $R=F$, 
$C_f = (N_c^2-1)/(2N_c)$ and $T_f=1/2$.
\bibitem{nfintegral}
Here, $N_f$ is formally extended to the nonnegative real numbers, with the
understanding that the physical values are nonnegative integers.
\bibitem{nfcr}
For discussions of the intensive lattice program to determine $N_{f,cr}$ for a
given $G$ and $R$, see \cite{lgtreviews}.  We assume $N_f > N_{f,cr}$ here, to
be in the NACP.  For several theories, there is not yet a consensus among
lattice groups on the respective values of $N_{f,cr}$.
\bibitem{convergence}
%
For all $R$ cases considered, we have studied the ratios 
$\kappa_{p-1}/\kappa_p$ and $|d_{p-1}/d_p|$ to 
get estimates of the rapidity of convergence of the series 
Eqs. (\ref{gamma_ir_Deltaseries}) and (\ref{betaprime_ir_Deltaseries}).
These suggest that for $R=F$, $\gamma_{IR,\Delta_f^4}$ and
$\beta'_{IR,\Delta_f^5}$ should be reasonably accurate over a substantial part
of the NACP. The convergence may be slower for $R=adj$ and $R=S$, with the
narrower intervals $I_{IRZ}$ in these cases.
\end{thebibliography}
\end{document}